\documentclass[floatfix,superscriptaddress,a4paper,
               nofootinbib,preprint]{revtex4}

\usepackage{epsfig}
\usepackage{latexsym}
\usepackage{xspace}

\usepackage{inputenc}
\usepackage{indentfirst}
\usepackage{enumerate}
\usepackage{color}

\usepackage{amsmath}
\usepackage{amssymb}
\usepackage[english]{babel}
\usepackage{url}
\usepackage{hyperref}

\graphicspath{{plots/}}
\include{abbreviations}

\begin{document}

\title{Scaling of factorial moments in cumulative variables}

\author{Subhasis Samanta 
}
\email{subhasis.samant@gmail.com}
\affiliation{Jan Kochanowski University, Kielce, Poland}

\author{Tobiasz Czopowicz}
\email{tobiasz.czopowicz@cern.ch }
\affiliation{Jan Kochanowski University, Kielce, Poland}
\affiliation{Warsaw University of Technology, Poland}

\author{Marek Gazdzicki}
\email{marek.gazdzicki@cern.ch}
\affiliation{Jan Kochanowski University, Kielce, Poland}
\affiliation{Geothe-University Frankfurt am Main, Germany}

\begin{abstract}

A search for power-law fluctuations within the framework of the intermittency method is ongoing to locate the critical point of the strongly interacting matter. In particular, experimental data on proton and pion production in heavy-ion collisions are analyzed in transverse-momentum, $p_T$, space.

In this regard, we have studied the dependence of the second scaled factorial moment $F_2$ of particle multiplicity distribution on the number of subdivisions of transverse momentum-interval used in the analysis. The study is performed using a simple model with a power-law two-particle correlation
function in $p_T$. We observe that $F_2$ values depend on the size and position of the $p_T$ interval. However, when we convert the non-uniform transverse-momentum distribution to uniform one using cumulative transformation, $F_2$  calculated in subdivisions of the cumulative $p_T$
becomes independent of the cumulative-$p_T$ interval.
The scaling behaviour of $F_2$ for the cumulative variable is observed. Moreover, $F_2$ follows a power law with the number of subdivisions of the cumulative-$p_T$ interval with the intermittency index close to the correlation function's exponent.

\end{abstract}

\pacs{....}
\keywords{...}
\maketitle

\section{Introduction}
\label{sec:introduction}

One of the central issues in high-energy heavy-ion physics is to locate the critical point (CP) in the phase diagram of the strongly interacting matter. Theoretical studies suggest that there should be a smooth crossover transition between the hadronic phase and quark-gluon plasma (QGP) at high temperature $T$ and zero baryon chemical potential $\mu_B$ ~\cite{Aoki:2006we}. A first-order phase transition is expected at small $T$ and large $\mu_B$~\cite{Asakawa:1989bq}. The CP is when the first-order phase transition line ends and has properties of second-order phase transition. Experiments at the CERN SPS and BNL RHIC are ongoing to search for the CP~\cite{Davis:2019mlt, STAR:2020tga, Luo:2017faz}. In future, there will be similar programmes in the Nuclotron-based Ion Collider Facility (NICA) at JINR, Dubna and the Compressed Baryonic
Matter (CBM) experiment at the Facility for Antiproton and Ion Research (FAIR) at GSI, Darmstadt \cite{Ablyazimov:2017guv}.
To find the CP, a search for power-law fluctuations within the framework of an intermittency method was proposed~\cite{Bialas:1985jb, Bialas:1988wc, Satz:1989vj, Gupta:1990bi}. 
Experimental data on proton and pion multiplicity fluctuations are analyzed in transverse
momentum, $p_T$, space~\cite{Anticic:2012xb,Anticic:2009pe,Davis:2020fcy}.

In this paper, we will show that when using the so-called cumulative transverse momentum for intermittency studies, the CP signal is independent of the size and location of the $p_T$ interval used in the analysis.
The paper is organized as follows. In Sec.~,\ref{sec:intermittency} we will briefly discuss the intermittency method. In Sec.~,\ref{sec:results} we present the results and finally, in Sec.~,\ref{sec:summary} we summarize and conclude the paper.

\subsection{Intermittency and scaling behaviour near CP}
\label{sec:intermittency}
In the intermittency method, observable of interest is the scaled factorial moment of particle multiplicity distribution. Its dependence on the number of subdivisions of momentum space is studied. 
The $q$-th order scaled factorial moment is defined as follows:
\begin{equation}
F_q (M) = \frac{ \langle \frac{1}{M^D} \sum_{i=1}^{M^D} n_i (n_i-1)...(n_i-q+1) \rangle }{ \langle \frac{1}{M^D} \sum_{i=1}^{M^D} n_i \rangle ^q },
\end{equation}
where $D$ is the dimension of the momentum space, $M^D$ is
the number of equally sized subdivisions of 
the $D-$dimensional space, $n_i$ is the number of particles in $i-$th subdivision. Here $\langle .. \rangle$ denotes the average is over the events.
At the critical point, the scaled factorial moment is expected to show a power-law dependence 
\begin{equation}
F_q (M) = (M^{D})^{\phi_q}
\end{equation}
when $M^D$ is sufficiently large.
The corresponding intermittency indices $\phi_q$ are expected
to follow the relation
\begin{equation}
 \phi_q = \frac{(q-1)}{D} d_q,
\end{equation}
where $d_q$ is the anomalous fractal dimension~\cite{DeWolf:1995nyp}.

The present work aims to study the dependence of the intermittency CP signal on the momentum interval selected for the analysis. For this purpose, a simple model is developed with a power-law two-particle correlation function.
For the simplicity we will consider 
a one dimensional momentum space ($D =1$)
and, thus, a two-particle distribution function of the form:
\begin{equation}\label{eq:corr}
 \rho(X_1,X_2) =  \frac{\rho({X_1}) \rho({X_2})}{ \left| X_1-X_2 \right| ^{\phi_2} + \epsilon}~,
\end{equation}
where $\rho(X)$ is a single particle distribution in $X$.
Furthermore we assume that
$X$ stands for dimensionless transverse momentum  $X \equiv p_T$/(1~GeV/$c$) and
it is distributed as
\begin{equation}\label{eq:rho}
 \rho(X) = C X e^{-6X},
\end{equation}
where $C$ is a normalization constant. 
The $\phi_2$ in Eq.~\ref{eq:corr} is the correlation function exponent. Throughout the analysis, we have taken $\phi_2 = 0.8$ as suggested by a CP model for protons~\cite{Antoniou:2006zb}.
At the denominator of Eq.~\ref{eq:corr} a small number $\epsilon$ ($= 10^{-6}$)
is added to avoid singularity for $X_1 = X_2$. 

\begin{figure}
\centering 
 \begin{center}
\includegraphics[width=0.45\textwidth]{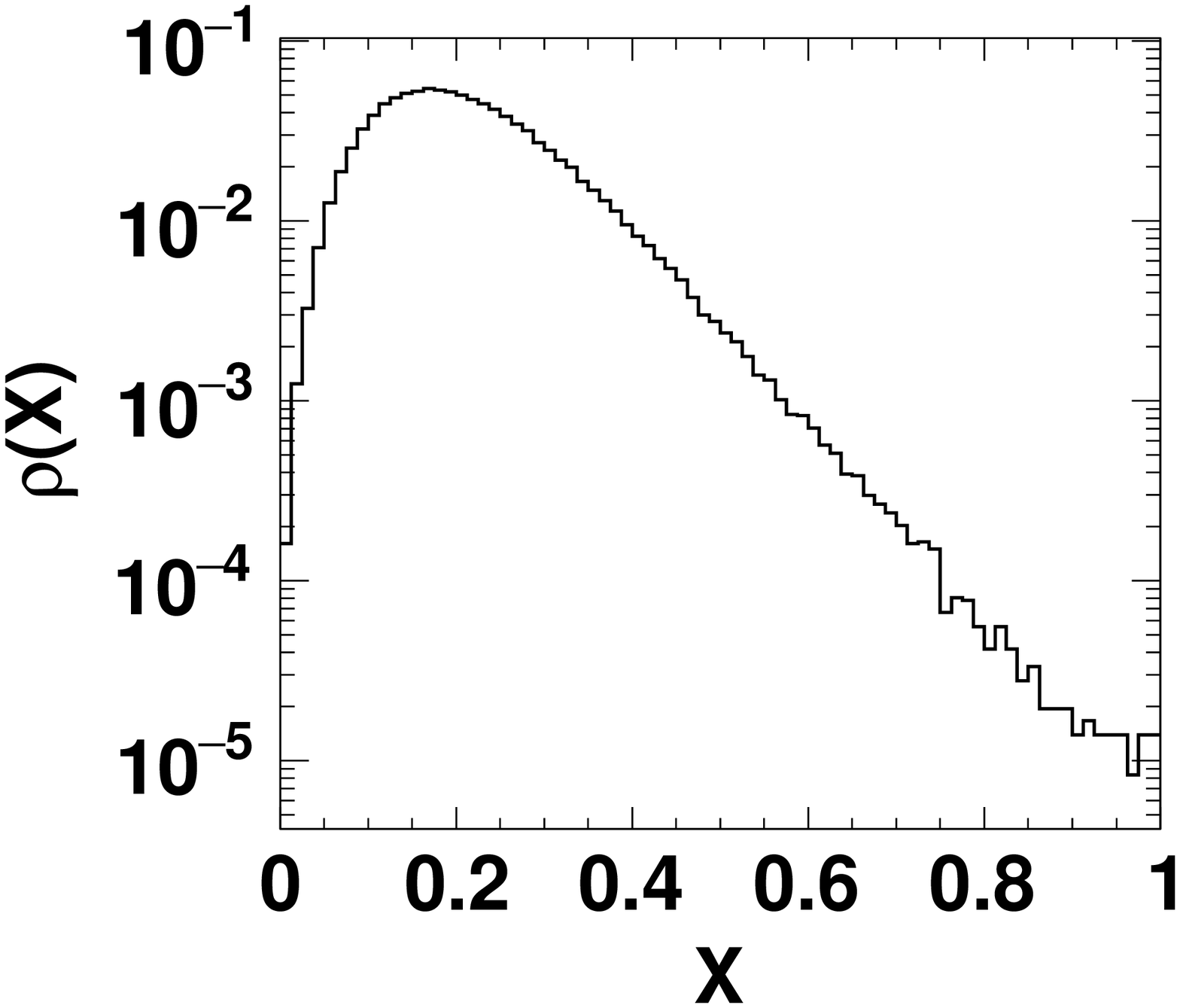}
\includegraphics[width=0.45\textwidth]{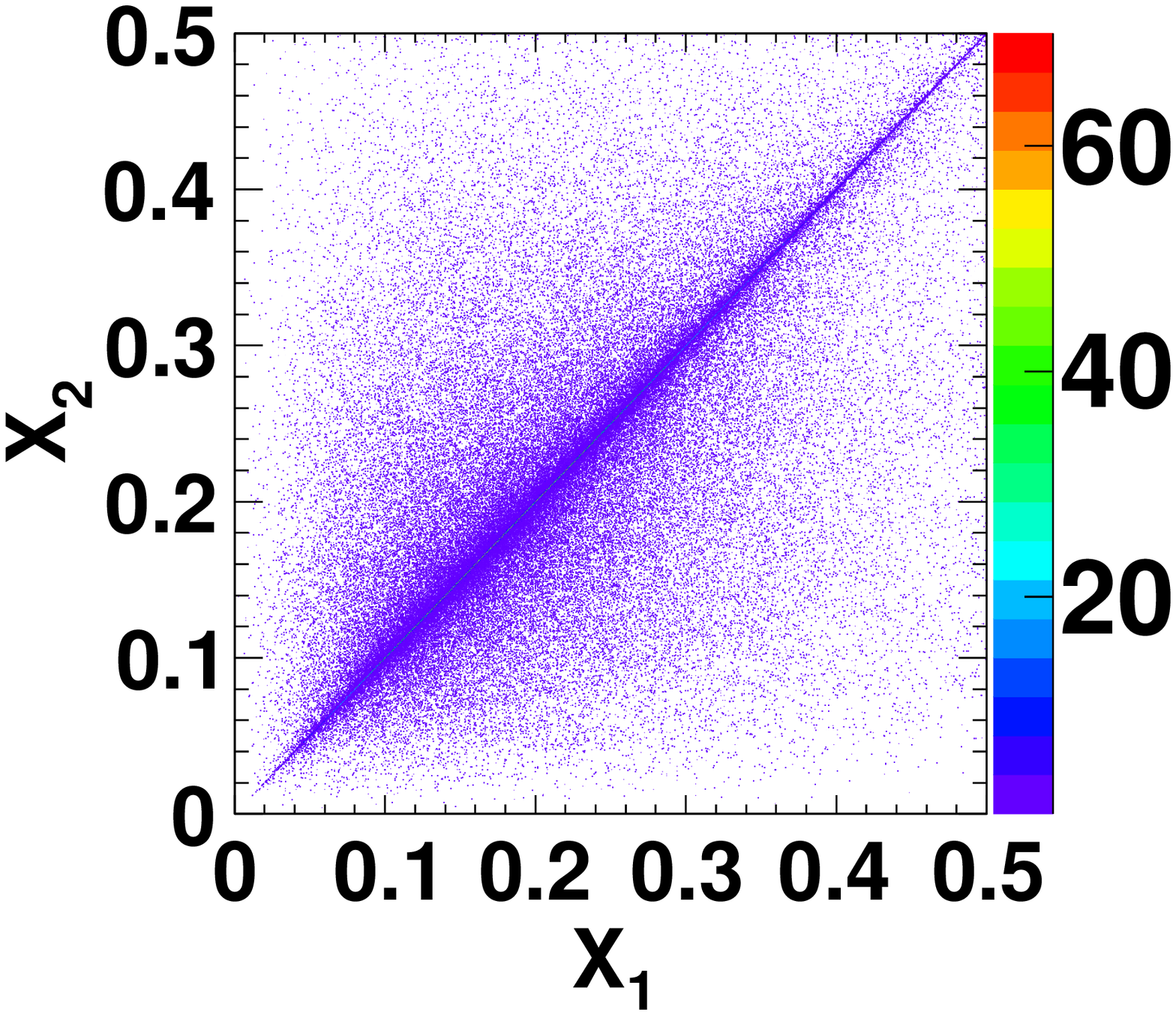}
\end{center}
\caption{
Distribution of $X$ given by Eq.~\ref{eq:rho}~(\textit{left}) and two particle  distribution in $X_1 - X_2$ plane generated according to Eq.~\ref{eq:corr}~(\textit{right}).}
\label{fig:X_dis}
\end{figure}

In the left panel of Fig.~,\ref{fig:X_dis} we show a distribution of $X$, which follows Eq.~\ref{eq:rho}. Here, we have used 180 thousand events with one correlated pair per event. Particles are generated within the $X$ range 0-1. Two particle distribution function following Eq.~\ref{eq:corr} is shown in the right panel of Fig.~\ref{fig:X_dis}. Due to the strong correlation, most particles are produced near the line $X_1 = X_2$.

The 
second scaled factorial moment can be expressed by a mean number of particle pairs as 
\begin{equation}
F_2 (M) = \frac{2M}{ \langle N \rangle^2 } \langle N_{pp}(M) \rangle,
\end{equation}
where $N$ and $N_{pp}$ are respectively multiplicity and the total number of particle pairs in all
$M$ subdivisions for a given event.

\begin{figure}
\centering
\begin{center}
\includegraphics[width=0.45\textwidth]{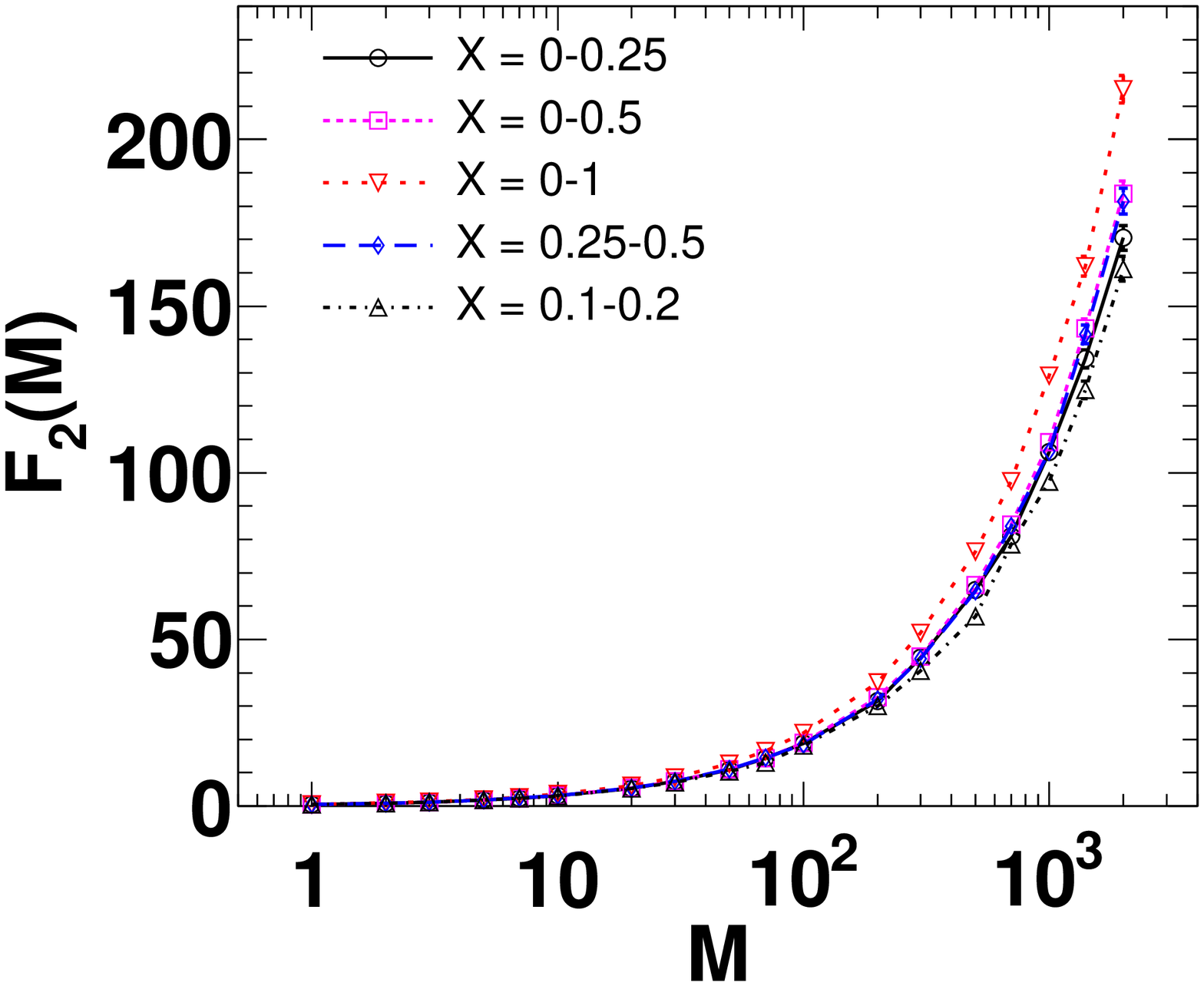}
\includegraphics[width=0.45\textwidth]{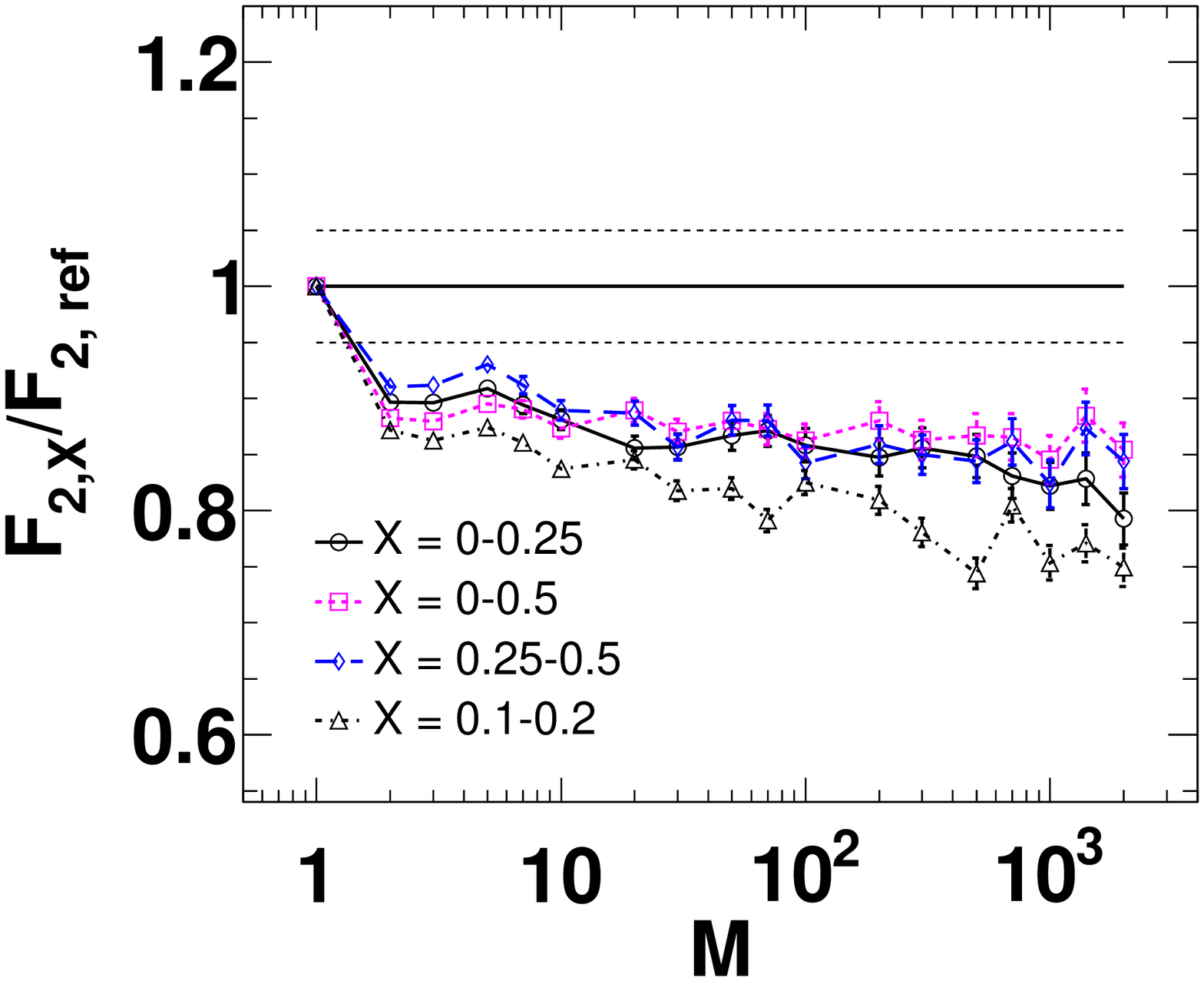}
\end{center}
\caption{\textit{Left:} Dependence of $F_2(M)$ on $M$ for different intervals of transverse momentum $X$. Here $F_2(M)$ is calculated in subdivisions of $X$. \textit{Right:} Ratio of $F_2(M)$ calculated for a given interval of $X$ to a reference $F_2(M)$ calculated for the $X$ interval (0, 1).}
\label{fig:F2_nonuniform}
\end{figure}

\section{Results}\label{sec:results}

In the left panel of Fig.~,\ref{fig:F2_nonuniform} we have shown $F_2$ as $M$ for different $X$ intervals. Here $F_2$ is calculated in subdivisions of a given interval of $X$. Data is generated independently for each $X$ interval. The first three data sets lower limit of $X$ is kept fixed at zero while the upper limits are 0.25, 0.5 and 1, respectively. For other two data sets $X$ intervals are (0.25, 0.5) and (0.1, 0.2).
$F_2$ at each $M$ is calculated from 10000 independent events, where each event consists of one correlated pair. $F_2$ is calculated at 18 different $M$. For each $X$ interval, we have used 180 thousand independent events.
Statistical uncertainty  of $F_2$ is calculated using the standard error propagation technique. At $M =1$, $F_2$ is 0.5 for each set since the number of pairs is one in each event. Further, $F_2$ increases with the increase of $M$. For $M>1$, values of $F_2$'s are different for different $X$ intervals. 
To quantify these differences, we have also calculated the ratio of $F_2$ at a particular $X$ interval to a reference $F_2$ obtained in the interval (0, 1). Note that different event sets are used to calculated the numerator and the denominator of this ratio to avoid correlation between results.
Ratios are shown in the right panel of Fig.~\ref{fig:F2_nonuniform}. Other than a line at 1, two more lines at 1.05 and 0.95 are drawn to indicate $\pm 5$ \% deviation from the reference $F_2(M)$. We observe that for all $X$ intervals, the deviation is greater than 5\% for $M>1$. Note that the distribution of $X$ considered in this work is highly non-uniform, shown 
in Fig.~\ref{fig:X_dis}~(\textit{left}).

Next we have analysed dependence of $F_2$ on $M$ for subdivisions in cumulative $X$ \cite{Bialas:1990dk} defined as:

\begin{equation}\label{eq:Q_X}
 Q_X = \frac{\int_{a}^{X} \rho(X) dX}{\int_{a}^{b} \rho(X) dX}~,
\end{equation}
where $a$ and $b$ are respectively the lower and upper limits of interval in $X$.

\begin{figure}
\centering
 \begin{center}
\includegraphics[width=0.45\textwidth]{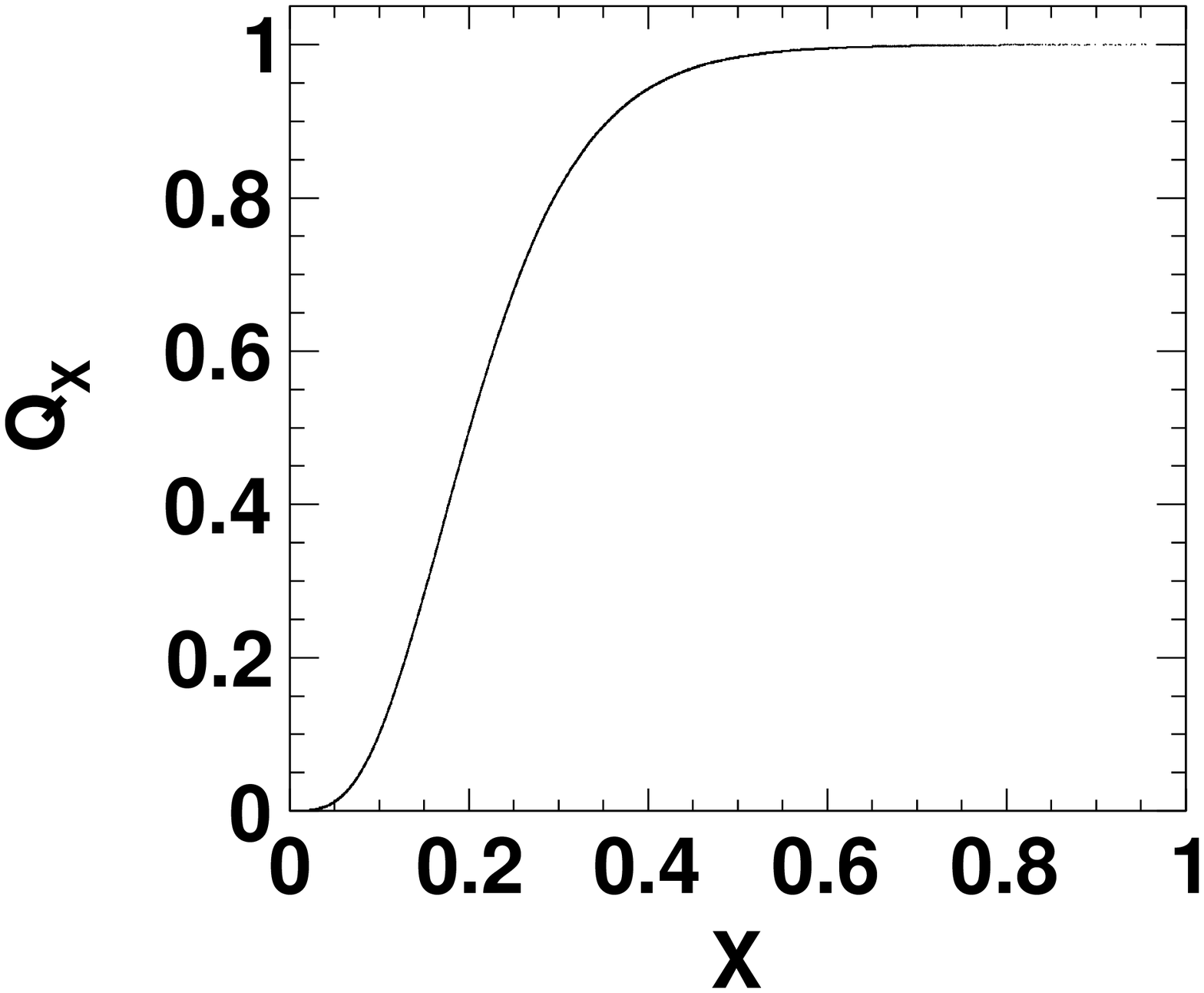}
\includegraphics[width=0.45\textwidth]{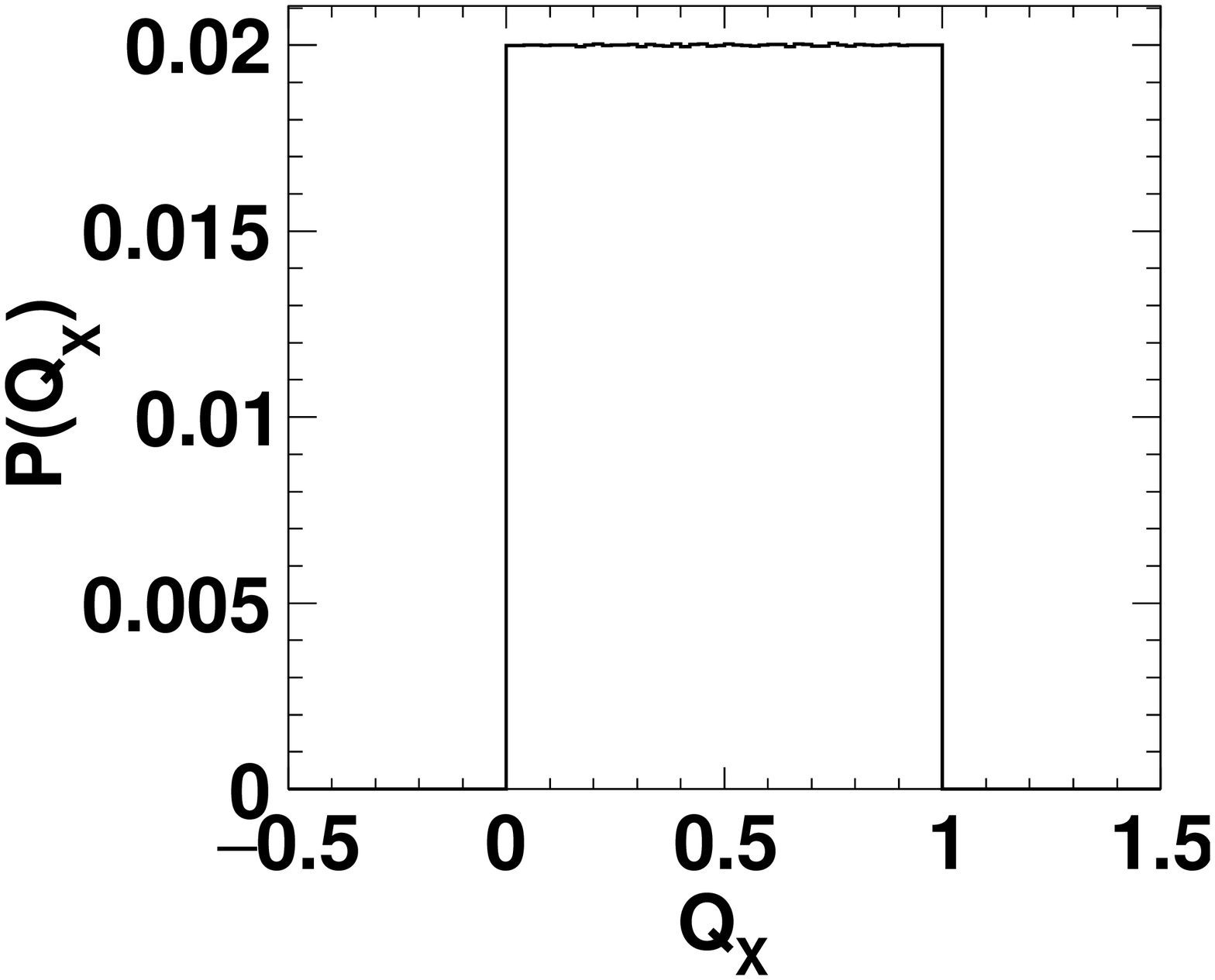}
\end{center}
\caption{\textit{Left:} Dependence of $Q_X$ on $X$. \textit{Right:} Uniform distribution of $Q_X$.}
\label{fig:QX_dis}
\end{figure}

According to Eq.~\ref{eq:Q_X}, $Q_X$ varies between zero to one and its distribution is uniform. The dependence of $Q_X$ on $X$ is shown in 
Fig.~\ref{fig:QX_dis}~(\textit{left}). As cross-check of the transformation
Fig.~\ref{fig:QX_dis}~(\textit{right}) shows the uniform distribution of $Q_X$.

\begin{figure}
\centering
 \begin{center}
\vspace{-0.2cm}
\includegraphics[width=0.45\textwidth]{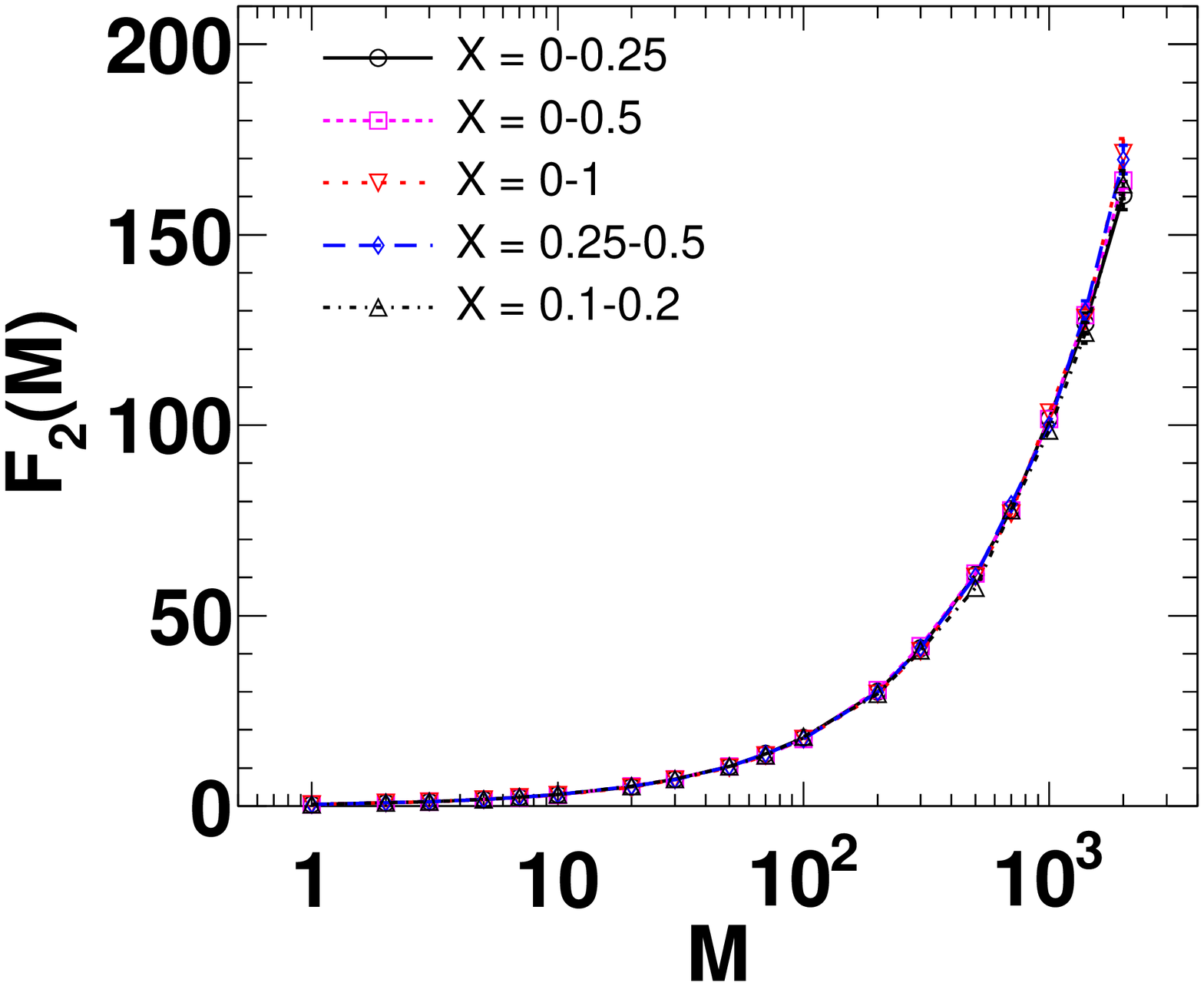}
\includegraphics[width=0.45\textwidth]{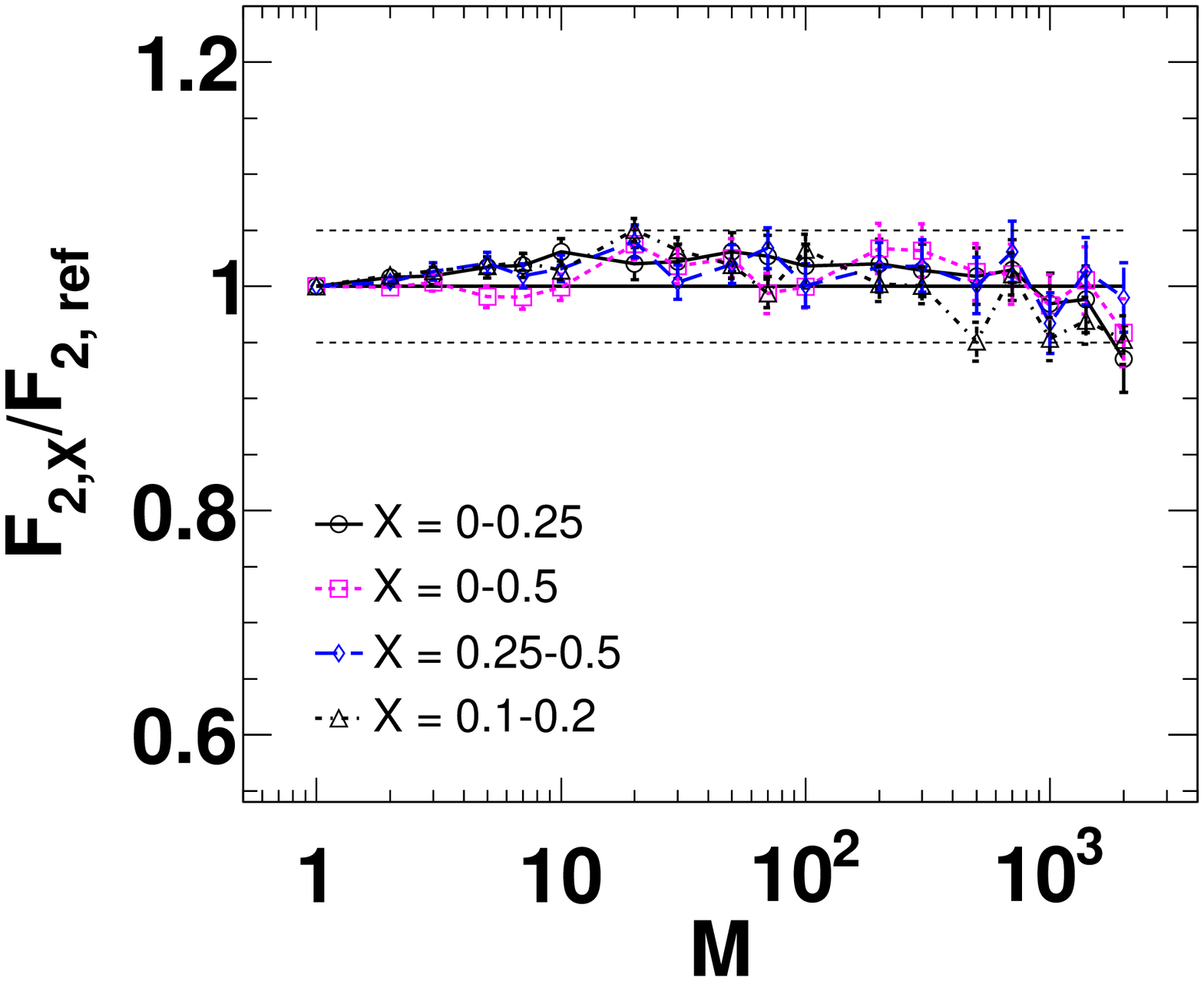}
\end{center}
\caption{Same as in Fig.~\ref{fig:F2_nonuniform}
but for subdivisions of cumulative transverse momentum $Q_X$.}
\label{fig:F2_uniform}
\end{figure}

We have calculated $F_2$ for subdivisions in $Q_X$, and the two-particle distribution is as previously given in $X_1$ and $X_2$. The left panel of Fig.~\ref{fig:F2_uniform} shows the dependency of $F_2$ on $M$ for different $X$ intervals. We observe that $F_2$'s are almost the same for all tested $X$ intervals. The ratio of $F_2$ in a particular $X$ interval to a reference $F_2$ are shown in 
Fig.~\ref{fig:F2_uniform}~(\textit{right}). We observe that the ratios for different $X$ intervals are consistent with each other and with the reference line within statistical uncertainties. Thus $F_2(M)$ calculated in the cumulative variable is independent of interval in $X$ selected for the analysis - the scaling of $F_2(M)$ in $M$ is observed. One may expect a scaling violation at larger $M$ due to a non-zero value of the parameter $\epsilon$ in Eq.~\ref{eq:corr} needed to avoid singularity at $X_1 = X_2$.

\subsection{Extraction of two-particle distribution function exponent}

\begin{figure}
\centering
 \begin{center}
\includegraphics[width=0.45\textwidth]{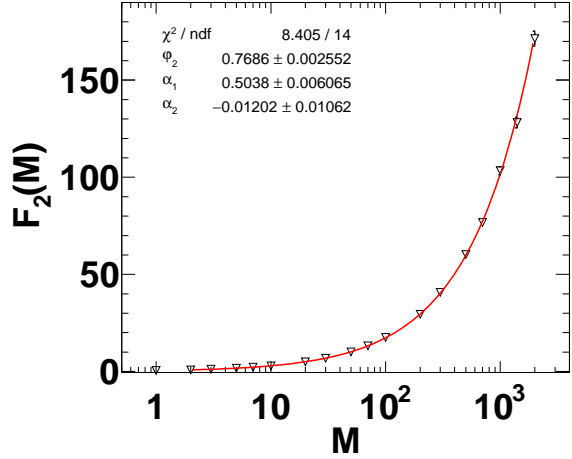}
\end{center}
\caption{Dependence of $F_2(M)$ on $M$ for subdivisions in cumulative variable for the $X$ interval (0, 1) fitted by the power-law function, Eq.~\ref{eq:fit_fn}.}
\label{fig:F2_fitting}
\end{figure}

Finally we have fitted the $F_2$  dependence on $M$ for the $X$ interval (0, 1) with a power-law function:
\begin{equation}\label{eq:fit_fn}
f(M) = \alpha_1 M^{\varphi_2} + \alpha_2
\end{equation}
where $\alpha_1, \alpha_2$ and $\varphi_2$ are parameters.For the fitting purpose we use 17 points within the $M$ range 2-2000. The fitted data and function
are shown in Fig.~\ref{fig:F2_fitting}. The $\chi^2$/ndf of the fitting is $\approx 0.6$.
The extracted power $\varphi_2 \approx 0.77$ (intermittency index) is close to the exponent $0.8$, used to generate correlated pairs.

\section{Summary and conclusion}
\label{sec:summary}
This work is motivated by experimental searches for the critical point of the strongly interacting matter in heavy-ion collisions. We have introduced a model with a power-law two-particle distribution function in dimensionless transverse momentum $X$. We have studied the behaviour of the second scaled factorial moment $F_2$ as a function of the number of subdivisions $M$ of $X$. The non-uniform distribution of $X$ $F_2(M)$ depends on the interval in $X$ selected for the analysis. However, once we convert the non-uniform distribution to a uniform distribution using cumulative variable $Q_X$ and calculated $F_2$ in subdivisions of $Q_X$, 
The scaling behaviour of $F_2(M)$ in $M$ is observed. 
Further, we have fitted dependence of $F_2(M)$ on $M$ with the power-law function $\alpha_1 M ^{\varphi_2} + \alpha_2$ and observed that the resulting intermittency index $\varphi_2$ is close to the exponent $\phi_2$ of the two-particle distribution function.

In a real experiment, $F_2$ will be affected by many biasing effects like detector efficiency, resolution etc. In Ref.~\cite{Samanta:2021usk}, we have shown that  $F_2$ is significantly modified by the effect of detector resolution. Usually, the particle detection efficiency is less than 100\%, and it may depend on transverse momentum~\cite{NA61:2014lfx, STAR:2020tga}. In future, we plan to study the effect of detector efficiency on $F_2$.

\begin{acknowledgments} 
This work was supported by 
the Polish National Agency for Academic Exchange through Ulam Scholarship with Agreement
No: PPN/ULM/2019/1/00093/U/00001,
the Polish National Science Centre grant 2018/30/A/ST2/00226 and the German Research Foundation grant GA1480\slash 8-1.

\end{acknowledgments}

\bibliographystyle{ieeetr}
\bibliography{references}
\end{document}